# 'Just a Normal Day in the Metaverse' – Distraction Conflicts of Knowledge Work in Virtual Environments

## Research-in-progress

### Julian Marx
Department of Computer Science and Applied Cognitive Science
University of Duisburg-Essen
Duisburg, Germany
Email: julian.marx@uni-due.de

### Jonas Rieskamp
Department of Information Systems
Paderborn University
Paderborn, Germany
Email: jonas.rieskamp@upb.de

### Milad Mirbabaie
Department of Information Systems
Paderborn University
Paderborn, Germany
Email: milad.mirbabaie@upb.de

## Abstract

The changing nature of knowledge work creates demands for emerging technologies as enablers for workplace innovation. One emerging technology to potentially remedy drawbacks of remote work arrangements are meta-verses that merge physical reality with digital virtuality. In the literature, such innovations in the knowledge work sector have been primarily examined against the backdrop of collaboration as a dependent variable. In this paper, however, we investigate knowledge work in meta-verses from a distraction-conflict perspective because independent, uninterrupted activities are as much characteristic of knowledge work as collaboration. Preliminary findings show that knowledge workers in meta-verses experience arousal from the 1) presence, appearance, and behaviour of other avatars, 2) realism, novelty, and affordances of the virtual environment, and 3) technological friction and navigation. This work has the theoretical implication that distraction-conflict theory must be extended to incorporate additional sources of arousal when applied to the context of knowledge work in meta-verses.

**Keywords** Metaverse, knowledge work, distraction-conflict theory, virtual environments.

# 1 Introduction

Emerging technologies fundamentally change the way we work (Wang et al. 2020). Fuelled by the COVID-19 pandemic, this process has gained additional acceleration. For instance, knowledge workers, that is, *"workers whose input is knowledge resources to yield knowledge-based intellectual output"* (Kianto et al. 2019, p.179) partly replaced travelling, social gatherings, or 9-5 office work with technology use. Thus, the interest in and use of technology for work increased substantially (Baptista et al. 2020). However, the resulting demand for remote working opportunities, globalisation, and cross-national collaboration affords suitable digital workplaces beyond existing solutions (Hafermalz and Riemer 2020; Mirbabaie et al. 2020).

As working remotely has become integral for many organisations, knowledge workers are particularly affected by change as their work can be considered location independent (Marx et al. 2021). In this respect, organisations begin to consider metaverse applications to overcome physical limitations of contemporary knowledge (Purdy 2022). The metaverse, or meta-verses, is an umbrella term for virtual environments that can be accessed via virtual reality (VR) or augmented reality (AR) devices. This creates a *"perpetual and persistent multiuser environment merging physical reality with digital virtuality"* (Mystakidis 2022, p.486). In these virtual environments, users are typically represented by avatars to interact and communicate with each other (Park and Kim 2022).

The default narrative of popular culture dealing with or research exploring meta-verses, we argue, is the one of meta-verses providing innovative meeting-space, increasing social presence, and facilitating collaboration (Brünker et al. 2022). With this paper, we problematise this view as it largely neglects the very essence of knowledge work, that is, individual, uninterrupted, and focused solo work. If meta-verses are to provide an alternative remote workplace, they must enable high performance work to maintain productivity (Lee et al. 2022; Xi et al. 2022). If collaboration is not the only dependent variable for researching knowledge work in meta-verses, what else is there except for performance? This paper argues from a theoretical angle that puts the elimination of distraction at the centre of this problem.

In their seminal work on distraction-conflict theory, Groff et al. (1983) found that visual and auditory distractions can increase the performance of an individual performing a simple task, whereas the same distraction tends to decrease the performance of complex tasks. Distractions arise due to internal or external arousals, which in turn facilitate or inhibit the performance of a task (Baron 1986; Nicholson et al. 2005). If meta-verses are implemented as remote work environments in professional contexts, it is imperative for Information Systems research to understand to what extend the theoretical mechanisms of distraction-conflict theory apply to the realm of this emerging technology. To contrast the literature on technology-enabled collaboration (e.g., Waizenegger et al. 2020), this paper proposes the investigation of metaverse technology from a distraction-conflict rather than collaborative view on knowledge work. In doing so, we go beyond the question about how technology facilitates collaboration but consider how technology such as meta-verses can enable distraction-free knowledge work that is impaired through remote work settings. Thus, we pose the following research question:

**RQ:** *What sources of arousal do knowledge workers identify in a metaverse environment?*

By answering this research question, we lay the groundwork for theoretical advancements as we (1) shed light on the phenomenological experiences of knowledge workers using meta-verses, (2) test the boundaries of distraction-conflict theory with respect to meta-verses, and (3) counterbalance the prevalence of empirical evidence of technologically enabled knowledge work revolving around collaboration. To answer our research question, we conducted qualitative experiments following the three phases of *pre-test*, *application of stimulus*, and *post-test* according to Robinson and Mendelson (2012). In the post-test phase, we conducted interviews with our participants to gain insights in their experience. The *application of stimulus* corresponds to the participants performing knowledge work tasks in a VR metaverse application. Specifically, the participants perform a simple and a complex task under the influence of different levels of arousal caused by the presence of other avatars and virtual objects.

# 2 Background

## 2.1 Knowledge Work in Meta-verses

The term *metaverse* relates to a continual environment that merges physical reality with digital virtuality (Mystakidis 2022). We use the plural *meta-verses* as many different technologies and platforms emerge that fit this definition but lack interoperability. In meta-verses, individuals can interact with others, do business, or forge social connections through virtual avatars (Duan et al. 2021;

Park and Catrambone 2007). Meta-verses have four characteristic features: realism, ubiquity, interoperability, and scalability. Realism means that the virtual environment provides the possibility for users to feel psychologically and emotionally immersed. Ubiquity refers to users connecting to meta-verses with everyday devices such as tablets, smartphones, or PCs. Interoperability signifies the ability of computer systems or software to exchange and make use of information for all users. The scalability of meta-verses is reflected in networks that are powerful enough to host many users at the same time and do not face recurrent technical problems (Dionisio et al. 2013; Mostajeran et al. 2022). The most prominent application of metaverse technology to date can be found in the video game industry. Here, meta-verses constitute micro-economies, which allow users to play, earn digital tokens, work, buy and sell virtual lands, learn, and communicate with others (Park and Kim 2022).

Nowadays, as globalisation has permeated the business world, including knowledge work sectors, organisations face increased costs of cooperation over large distances. Emerging technologies such as meta-verses can decrease those costs. For example, in 2020, UC Berkeley held its graduation ceremony in the video game Minecraft (Duan et al. 2021). Furthermore, the scholarly literature provides several examples in which meta-verses have been subjects of scrutiny in different contexts of work. For example, Liu and Yu (2018) conducted an experiment in a virtual environment to test whether a virtually present co-actor can elicit a social facilitation effect in visual search tasks. The results of this study show that virtually present co-actors can shorten the response time of participants when they are completing easy tasks but lengthen it when they are completing difficult tasks. In another paper, Miller et al. (2019) arranged three scenarios to assess the social effects of AR use. The results revealed social facilitation and deterrence effects in easy tasks and detrimental effects in an arduous task in the presence of the virtual agent. Jeffri and Rambli (2021) reviewed literature on AR systems and their effects on mental workload and performance. They found that if the mental workloads are positive, the effects on the task performance are more likely to be positive as well and wise versa.

## 2.2 Distraction-Conflict Theory

The distraction-conflict theory was described for the first time by Groff et al (1983). The theory emphasizes that, rather than the mere attendance of others, it is the contradiction between giving attention to a person and giving attention to a task that affects performance. It means that the presence of a second person or object, while another person is doing a job, can have a positive or negative effect on the performance of that worker. Along with this, an individual's performance on simple tasks is facilitated by arousal, whereas an individual's performance on complex tasks is hindered by the same arousal. These phenomena happen due to both audiences and competitors distracting subjects from the empirical task and as a result creating attentional conflict (Baron 1986). In simple words, this theory conveys that the presence of others can be an arousal. In turn, this arousal can increase the performance of the worker's job if it causes social facilitation. If it causes social inhibition, it decreases the performance. This mechanism is shown in Figure 1.

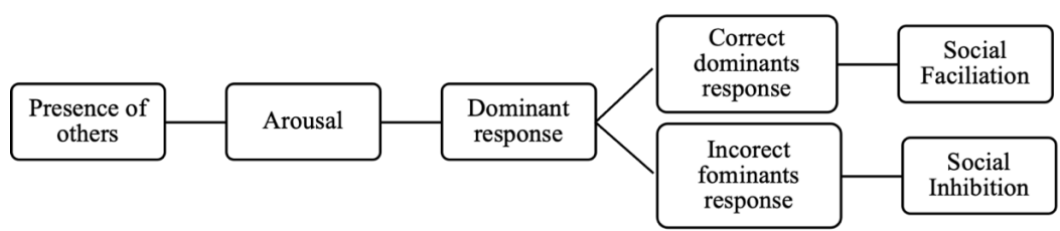

*Figure 1: Model of the distraction-conflict theory, based on Baron (1986).*

Distraction can have several dimensions: social vs. non-social, external stimulus vs. internal thought, and imposed by a second party or created by individuals themselves (Nicholson et al. 2005). Several previous studies have shown that visual and auditory distractions cause the facilitation of simple task performance and impair complex task performance (Groff et al. 1983). If the task requires a high amount of cognitive effort, the presence in an environment where distractions are present may decrease the performance of that task (Nicholson et al. 2005). Knowledge work in meta-verses, we argue, may be subject to this distraction-conflict as virtual environments provide an abundance of possible distractions. Extant research already showed that social media use at work can cause distraction, leading to negative consequences (Brooks et al. 2017). Similarly, distraction by mobile devices is negatively associated with the users' mental well-being (Chu et al. 2021). In the realm of virtual reality, it has been suggested to limit the interactivity and vividness of the environment to avoid information overload caused by strong distractions (Chen et al. 2022).

## 3 Research Design

To answer our research question, we collected data by conducting qualitative experiments. To this end, we acquired eight full-time knowledge workers from the academic sector as participants. They were between 23 and 29 years old, three of them were female, and no prior metaverse experience was needed. The research design follows the three phases *pre-test*, *application of stimulus*, and *post-test* according to Robinson and Mendelson (2012). In the first phase, participants were briefed regarding the procedure of the experiment. In the second phase, the participants had to perform a task in a metaverse application which differed in complexity and the level of presence of other avatars and virtual objects. In line with distraction conflict theory, a simple and a complex task had to be performed (cf. Table 1).

| Difficulty | Task |
|---|---|
| simple | Online search for an image to use in a presentation representing the topic of "Carbon Dioxide and its Effects on the Environment". |
| complex | Online search for a scientific publication explaining the topic of "Carbon Dioxide and its Effects on the Environment". |

*Table 1. Task description of the qualitative experiment*

The experiment was conducted in the metaverse application "AltSpace VR" while wearing Oculus Quest 1 VR goggles. Using this application provides the opportunity of a private room in which the participant can work on the task without anybody being nearby. Further, the software provides a "campfire area", which is publicly accessible. Hence, it is suitable for the conduct of the task under the influence of arousals caused by the presence of other avatars.

The task complexity levels and locations in which the tasks were conducted yielded four different settings: (1) simple task in the private room, (2) complex task in the private room, (3) simple task in the public room, and (4) complex task in the public room. In the *post-test* phase, we conducted semi-structured interviews to examine the participants' experience of performing the tasks. The questions intended to capture the participants' perception of arousal during, and difficulty of the tasks performed. Thereby, the answers inform us about the applicability of the distraction-conflict theory. The interviews were recorded and transcribed (Miles and Huberman 2002). We conducted the analysis of the resulting transcripts inductively according to Mayring (2014).

## 4 Preliminary Results

In this section, we report preliminary results of our analysis pertaining to the initially articulated research question. The analysis of eight interview transcripts revealed three inductive codes that represent different sources of arousal the participants identified. These categories are presented in Table 2 alongside exemplary statements and sentiments.

### 4.1 Presence, appearance, and behaviour of other avatars

Most participants reported that they felt aroused by the presence of other persons. This is in line with what we could expect from distraction-conflict theory.

> *"I was distracted for a moment when you walked into the room, your avatar was a distraction as well, of course, but then I quickly went back to my task."* (P1)

In meta-verses, however, other individuals appear in the form of avatars that may represent the actual appearance of a person or is entirely fictional. For the participants, it was not only the presence of other avatars that caused arousal but also their visual appearance.

> *"The avatars themselves were a distraction because some of them were designed in a funny way."* (P2)

> *"The main distraction was definitely the presence of other avatars, mainly their appearance and the sounds that come with them."* (P5)

Apart from the visual appearance, the behaviour of other avatars was a source of distraction, especially in the open environment ("campfire"). In addition, some participants reported about distracting sounds and voices stemming from the avatars. In this case, some participants decided to activate a mute-function.

> *"The other users and their avatars were a distraction, some flew through the air, others went really crazy when I observed other conversations."* (P4)

### 4.2 Realism, novelty, and affordances of the virtual environment

Other avatars were not the only source of distraction. Some participants stated that they were overwhelmed by the *novelty* and *affordances* waiting in the virtual environment, which could potentially steal the attention away from the tasks.

> *"I was unsure at the beginning because I couldn't manage the difficult task and had to try out some things more often."* (P1)

> *"It was very exciting to meet strangers in this way, it was a bit like going to the park and meeting on a blind date."* (P2)

The perceived realism of the virtual environment was a surprise for participants who were first-time users. As a source of distraction, this effect might wear off over time. However, technological advancement and updates of meta-verses might continually enhance the perceived level of *realism* and therefore, evoke feelings of arousal.

> *"When I put on the VR glasses, I was amazed because I couldn't see anything of my surroundings directly because I was in a parallel world. It felt so real."* (P1)

> *"I felt lost and didn't know what to do exactly."* (P8)

### 4.3 Technological friction and navigation

A third code that was salient in our data refers to the perceived technological shortcomings of the experience. Most participants reported that the controls were not very handy, and it took some time to get used to them. The performance of the tasks was often inhibited by friction caused by the controls or the limitation of the technology.

> *"In the beginning it was a bit complicated because the whole experience was unusual to me."* (P4)

P7, for example, was frustrated by the functionalities of the system and named this as the principal reason for not using it for regular work.

> *"Some things in the metaverse like using the browser and researching didn't work well. But I think if the Metaverse and the use of the programs there got better, I could see myself studying in the Metaverse."* (P7)

At the same time, however, most participants anticipate the technology might find its way into their lives once it has improved. The concerns relate to both software and hardware (for long periods of work, the VR headset feels heavy and disturbs the immersive experience, P2 reported).

| Code | Example Statement | Sentiments (Examples) |
|---|---|---|
| #1 Presence, appearance, and behaviour of other avatars | *"The main distraction was definitely the presence of other avatars, mainly their appearance and the sounds that come with them."* (P5) | "crazy", "funny-looking" |
| #2 Realism, novelty, and affordances of the virtual environment | *"When I put on the VR glasses, I was amazed because I couldn't see anything of my surroundings directly because I was in a parallel world. It felt so real."* (P1) | "unusual", "unsure", "exciting", "realistic" |
| #3 Technological friction and navigation | *"I think I would prefer the real world because some things in the metaverse like using the browser and researching didn't work well. But I think if the Metaverse and the use of the programs there got better, I could see myself studying in the Metaverse."* (P7) | "frustrating", "I felt lost", "more difficult at the moment" |

*Table 2. Codes for different sources of arousal when performing knowledge work in meta-verses*

## 5 Discussion and Next Steps

The arousal experienced by participants had three main sources: 1) presence, appearance, and behaviour of other avatars, 2) realism, novelty, and affordances of the virtual environment, and 3) technological friction and navigation. These preliminary results suggest that in the context of meta-verses, knowledge workers might encounter additional sources of arousal as compared to regular work settings. Distraction-conflict theory deals with the presence of others as the main source of arousal, and consequently, distraction (Baron 1986). In the context of meta-verses, however, we argue for broadening this view and suggest researchers to consider other independent variables when conducting research in the context of knowledge work. According to the distraction-conflict theory, the performance of simple tasks is enhanced with others present. In turn, the performance of complex tasks decreases in the same setting (Nicholson et al. 2005). A full version of this study is set out to derive further theoretical implications that refer to the core mechanism of distraction-conflict theory. Our first impression is that results are in line with the assumptions of the theory and other literature, suggesting that complex tasks are indeed harder to accomplish with other avatars present in meta-verses (Mostajeran et al. 2022). Moreover, the task complexity levels and locations yielded insights about different settings. Results from further experiments could be interpreted considering these four settings.

The present study was subject to limitations. The research team was confined to using "Altspace VR" as other metaverse applications such as "Horizon World" or "Horizon Workrooms" were not available at the time of the experiments outside of the US. Moreover, interoperability of many metaverse applications is limited. For example, some applications are being released for Occulus Quest 2, which are not compatible with Occulus Quest 1. Moreover, as discussed above, the low level of experience of many participants with the technology was a limitation during the experiments.

Next steps of this research will be the extension of the sample of participants. We aim for more breadth in terms of experience level, background, and demographics. Moreover, the interviews in the *post-test* phase will incorporate additional techniques from the phenomenological method to get participants to reflect upon their experience, which will result in rich qualitative data. The primary goal of an extended version of this work will not be to mimic existing research and finding evidence for the distraction-conflict mechanism in meta-verses. Rather, we aim to make a theoretical contribution by extending the theory with additional concepts that apply in virtual environments such as meta-verses. At the same time, this will allow us to test the boundaries of the theory and assess to what extend it is applicable to the *"exciting"*, *"crazy"*, and quite *"unusual"* metaverse experience.